\documentstyle[12pt]{article}

\newcommand{\sect}[1]{\setcounter{equation}{0}\section{#1}}

\def\bseq{\begin{subequation}}  
\def\eseq{\end{subequation}}
\def\bsea{\begin{subeqnarray}}  
\def\esea{\end{subeqnarray}}

\def\Dot#1{{\kern0.5pt
     {#1} \kern-5.05pt \raise5.8pt\hbox{$\textstyle.$}\kern
0.5pt}}

\newcommand{\beq}{\begin{equation}}
\newcommand{\eeq}{\end{equation}}
\newcommand{\bea}{\begin{eqnarray}}
\newcommand{\eea}{\end{eqnarray}}
\newcommand{\ena}{\end{eqnarray}}

\newcommand {\non}{\nonumber}

\renewcommand{\a}{\alpha}
\renewcommand{\b}{\beta}

\renewcommand{\d}{\delta}
\newcommand{\th}{\theta}

\newcommand{\pa}{\partial}

\newcommand{\g}{\gamma}

\newcommand{\e}{\epsilon}

\newcommand{\m}{\mu}

\newcommand{\n}{\nu}

\newcommand{\thb}{\bar{\theta}}

\def\Mb{\kern 2pt\mathchoice
            {
             \vbox{\hrule width10pt height 0.4pt depth 0pt
                 \kern 1.2pt\hbox{\kern -2pt$\displaystyle M$}}}
            {
                 \vbox{\hrule width10pt height 0.4pt depth 0pt
                 \kern 1.2pt\hbox{\kern -2pt$\textstyle M$}}}
            {
\vbox{\hrule width6pt height 0.4pt depth 0pt
                 \kern 1.0pt\hbox{\kern -2pt$\scriptstyle M$}}}
            {
                 \vbox{\hrule width5pt height 0.4pt depth 0pt
                 \kern 0.8pt\hbox{\kern -2pt$\scriptscriptstyle M$}}}}

\def\Sb{\kern 2pt\mathchoice
            {
                 \vbox{\hrule width6pt height 0.4pt depth 0pt
                 \kern 1.2pt\hbox{\kern -2pt$\displaystyle S$}}}
            {
                 \vbox{\hrule width6pt height 0.4pt depth 0pt
                 \kern 1.2pt\hbox{\kern -2pt$\textstyle S$}}}
            {
                 \vbox{\hrule width3.5pt height 0.4pt depth 0pt
                 \kern 1.0pt\hbox{\kern -2pt$\scriptstyle S$}}}
            {
                 \vbox{\hrule width3pt height 0.4pt depth 0pt
                 \kern 0.8pt\hbox{\kern -2pt$\scriptscriptstyle S$}}}}

\def\Rb{\kern 2pt\mathchoice
            {
                 \vbox{\hrule width5.5pt height 0.4pt depth 0pt
                 \kern 1.2pt\hbox{\kern -2.5pt$\displaystyle R$}}}
            {
                 \vbox{\hrule width5.5pt height 0.4pt depth 0pt
                 \kern 1.2pt\hbox{\kern -2.5pt$\textstyle R$}}}
            {
                 \vbox{\hrule width3.5pt height 0.4pt depth 0pt
                 \kern 1.0pt\hbox{\kern -2.2pt$\scriptstyle R$}}}
            {
                 \vbox{\hrule width3pt height 0.4pt depth 0pt
                 \kern 0.8pt\hbox{\kern -2.2pt$\scriptscriptstyle R$}}}}

  \def\pp{{\mathchoice
          {
              \kern 1pt%
              \raise 1pt
              \vbox{\hrule width5pt height0.4pt depth0pt
                    \kern -2pt
                    \hbox{\kern 2.3pt
                          \vrule width0.4pt height6pt depth0pt
                          }
                    \kern -2pt
                    \hrule width5pt height0.4pt depth0pt}%
                    \kern 1pt
           }
            {
              \kern 1pt%
              \raise 1pt
              \vbox{\hrule width4.3pt height0.4pt depth0pt
                    \kern -1.8pt
                    \hbox{\kern 1.95pt
                          \vrule width0.4pt height5.4pt depth0pt
                          }
                    \kern -1.8pt
                    \hrule width4.3pt height0.4pt depth0pt}%
                    \kern 1pt
            }
            {
              \kern 0.5pt%
              \raise 1pt
              \vbox{\hrule width4.0pt height0.3pt depth0pt
                    \kern -1.9pt  
                    \hbox{\kern 1.85pt
                          \vrule width0.3pt height5.7pt depth0pt
                          }
                    \kern -1.9pt
                    \hrule width4.0pt height0.3pt depth0pt}%
                    \kern 0.5pt
            }
            {
              \kern 0.5pt%
              \raise 1pt
              \vbox{\hrule width3.6pt height0.3pt depth0pt
                    \kern -1.5pt
                    \hbox{\kern 1.65pt
                          \vrule width0.3pt height4.5pt depth0pt
                          }
                    \kern -1.5pt
                    \hrule width3.6pt height0.3pt depth0pt}%
                    \kern 0.5pt
            }
        }}

  \def\mm{{\mathchoice
                       {
                             \kern 1pt
               \raise 1pt    \vbox{\hrule width5pt height0.4pt depth0pt
                                  \kern 2pt
                                  \hrule width5pt height0.4pt depth0pt}
                             \kern 1pt}
                       {
                            \kern 1pt
               \raise 1pt \vbox{\hrule width4.3pt height0.4pt depth0pt
                                  \kern 1.8pt
                                  \hrule width4.3pt height0.4pt depth0pt}
                             \kern 1pt}
                       {
                            \kern 0.5pt
               \raise 1pt
                            \vbox{\hrule width4.0pt height0.3pt depth0pt
                                  \kern 1.9pt
                                  \hrule width4.0pt height0.3pt depth0pt}
                            \kern 1pt}
                       {
                           \kern 0.5pt
             \raise 1pt  \vbox{\hrule width3.6pt height0.3pt depth0pt
                                  \kern 1.5pt
                                  \hrule width3.6pt height0.3pt depth0pt}
                           \kern 0.5pt}
                       }}

\def\pd{{\kern0.5pt
                   + \kern-5.05pt \raise5.8pt\hbox{$\textstyle.$}\kern
0.5pt}}

\def\pmd{{\kern0.5pt
                  \pm \kern-5.05pt
\raise6.3pt\hbox{$\textstyle.$}\kern1.5pt}}

\def\md{{\mathchoice
   {
      {{\kern 1pt - \kern-6.2pt \raise5pt\hbox{$\textstyle.$}\kern
1pt}}}
    {
      {{\kern 1pt - \kern-6.2pt \raise5pt\hbox{$\textstyle.$}\kern
1pt}}}
    {
      {\kern0.5pt - \kern-5.05pt
\raise3.4pt\hbox{$\textstyle.$}\kern0.5pt}}
    {
      {\kern0.5pt - \kern-5.05pt
\raise3.4pt\hbox{$\textstyle.$}\kern0.5pt}}}}

\newcommand{\ad}{{\dot{\alpha}}}
\newcommand{\bd}{{\dot{\beta}}}

\begin{document}

\begin{titlepage}
{\hbox to\hsize{April  2001 \hfill
{Bicocca--FT--01--11}}}
{\hbox to\hsize{${~}$ \hfill
{IFUM--686--FT}}}
\begin{center}
\vglue .05in
{\Large\bf Non(anti)commutative Superspace}
\footnote{Supported in
part by INFN, MURST and the European Commission RTN program
HPRN--CT--2000--00131, in which S.P. is associated to
the University of Padova and D.~K.~is associated to the University of
Torino. }\\[.45in]
Dietmar Klemm\footnote{dietmar.klemm@mi.infn.it}\\
{\it Dipartimento di Fisica, Universit\`a degli studi di Milano \\
and INFN, Sezione di Milano, via Celoria 16, 20133 Milano, Italy}
\\
[.2in]
Silvia Penati\footnote{silvia.penati@mib.infn.it}\\
{\it Dipartimento di Fisica dell'Universit\`a degli studi di
Milano-Bicocca\\
and INFN, Sezione di Milano, piazza della Scienza 3, I-20126 Milano,
Italy}
\\
[.2in]
Laura Tamassia\footnote{laura.tamassia@pv.infn.it}\\
{\it Dipartimento di Fisica Nucleare e Teorica dell'Universit\`a 
degli studi di Pavia and INFN, Sezione di Pavia,\\
via Ugo Bassi 6, I-27100 Pavia,
Italy}\\[.4in]

{\bf ABSTRACT}\\[.0010in]
\end{center}

We investigate the most general non(anti)commutative geometry in $N=1$
four dimensional superspace, invariant under the classical (i.e. undeformed)
supertranslation group. We find that a nontrivial
non(anti)commutative superspace geometry compatible with
supertranslations exists with non(anti)commutation parameters which may
depend on the spinorial coordinates. The algebra is in general
nonassociative. Imposing associativity
introduces additional constraints which however allow
for nontrivial commutation relations involving fermionic
coordinates. We obtain explicitly the first three terms 
of a series expansion in the deformation parameter 
for a possible associative $\star$--product.
We also consider the case of $N=2$ euclidean superspace where the
different conjugation relations among spinorial coordinates allow for
a more general supergeometry.

${~~~}$ \newline

\end{titlepage}

\sect{Introduction}

During the past two years a clear connection between string theory
and noncommutative geometry has emerged. In the presence of a
constant magnetic Neveu-Schwarz field, the low energy dynamics of D3 brane
excitations is described by noncommutative ${\cal N}=4$ super Yang-Mills
theory \cite{seibwitt}. The result by Seiberg and Witten followed
earlier work \cite{connes}, where it was found that noncommutative
geometry arises very naturally in the framework of M(atrix) theory.
Apart from the string theory context, noncommutative field theories
are interesting in their own right. This interest is motivated
by many intriguing features of field theories on noncommutative
spaces, like the UV/IR mixing \cite{minwalla} or the Morita equivalence
\cite{schwarz} between Yang-Mills theories on noncommutative tori.
Noncommutative field theories also play a role
in solid state physics, e.~g.~noncommutative Chern Simons theory was
recently proposed by Susskind \cite{susskind} to provide a description
of the fractional quantum Hall effect. Also in the physics of black
holes, noncommutativity of spacetime naturally emerges from
`t Hooft's S-matrix ansatz \cite{thooft} (Cf.~also \cite{li} for a recent
review). The hope is that this may eventually lead to a better
understanding of some of the puzzles of black hole physics, such
as the information loss paradox.

Up to now, the concept of noncommutativity has been limited
essentially to the bosonic coordinates, i.~e.~one has
\beq
[x^{\mu}, x^{\nu}] = i\Theta^{\mu\nu}
\label{bosonic}
\eeq
where $\Theta^{\mu\nu}(x)$ is antisymmetric. In view of the fact
that supersymmetry plays a fundamental role in string theory,
it seems natural and
compelling to ask what happens if we deform also the anticommutators
between fermionic coordinates of superspace\footnote{Non-anticommutative
structures in field theory and gravity have been studied in a
different context in \cite{Moffat:2000ac}. Furthermore,
nonvanishing anticommutators of fermionic coordinates have been
considered in \cite{Schwarz:pf} in the context of a
possible fermionic substructure of spacetime.}, or the commutators
between bosonic and fermionic coordinates. To investigate the most
general deformations compatible with supersymmetry is the main purpose
of this paper. First steps in this direction were undertaken
in \cite{kosinski}, where quantum deformations of the Poincar\'{e}
supergroup were considered, and in \cite{ferrara}, where it was
shown that in general chiral superfields are not closed under
star products that involve also deformations of fermionic coordinates.
Here we will be mainly concerned with the conditions imposed
on the possible deformations of superspace by requirements such
as covariance under {\em classical} translations and supertranslations,
Jacobi identities, associativity of the star product, and closure of
chiral superfields under the star product.

The paper is organized as follows: In Section 2 we
determine the most general deformation of four dimensional, $N=1$
Minkowski superspace that is covariant
under undeformed supertranslations, and discuss the deformation of the
supersymmetry algebra which follows. The result we obtain is a
supersymmetric nontrivial extension of the ``constant $\Theta$''
noncommutative bosonic geometry studied in a string theory context
\cite{seibwitt}. While our geometries remain flat in the bosonic sector,
they are curved along the fermionic directions.
In Section 3 we study
the restriction imposed on this general structure by the Jacobi identities,
i.~e.~by the requirement to have a super Poisson structure on superspace.
We will see that these additional constraints necessarily impose the
spinorial coordinates to be anticommuting, but allow for possible nontrivial
commutation relations among bosonic and fermionic coordinates. In this
case the standard supersymmetry algebra is restored.
It is then shown that the violation of the
Jacobi identities, implied by the deformation of the fermionic coordinates,
is equivalent to the nonvanishing of a super three-form field strength.
It might be interesting to study whether deformations of fermionic variables 
arise in superstring theory with backgrounds that involve nonvanishing
super p--form field strengths. In the following Section
we obtain the first three terms in a series expansion in the deformation 
parameter $\hbar$ for a possible noncommutative product of superfields.
Due to the 
violation of the Jacobi identities, this product will be in general 
nonassociative. In the
cases where the Jacobi identities are satisfied, we show that this
product is associative up to quadratic order. Finally, in the last 
Section we discuss possible
non(anti)commutative deformations for superspaces with euclidean signature.
In the simplest $N=2$ case, we find that deformations involving nontrivial
anticommutation relations among spinorial variables are in this case
allowed by the request of consistency with supercovariance and associativity.
We conclude with some final remarks.

\sect{Covariant non(anti)commutative geometry}

We consider a four dimensional $N=1$ Minkowski superspace.
The set of superspace coordinates are $Z^A = (x^{\a \ad}, \th^\a, \thb^\ad)$,
where $x^{\a \ad}$ are the four real bosonic coordinates and $\th^\a$,
$\thb^\ad$ two--component complex Weyl fermions.
The conjugation rule $\thb^\ad = (\th^\a)^{\dag}$ follows from the
requirement to have a four component Majorana fermion
(we use conventions of {\em Superspace} \cite{super}).

In the standard (anti)commutative superspace the algebra of the coordinates is
\bea
&& \{ \th^\a , \th^\b \} ~=~\{ \thb^\ad , \thb^\bd \} ~=~
\{ \th^\a , \thb^\ad \} ~=~ 0 \non\\
&& [ x^{\a \ad} , \th^\b ] ~=~ [ x^{\a \ad} , \thb^\bd ] ~=~ 0 \non\\
&& [ x^{\a \ad} , x^{\b \bd} ] ~=~ 0
\label{coord}
\eea
and it is trivially covariant under the superpoincar\'e group.
The subgroup of the classical (super)translations (spacetime translations and
supersymmetry transformations)
\bea
&& \th'^\a ~=~ \th^\a ~+~ \epsilon^\a \non \\
&& \thb'^\ad ~=~ \thb^\ad ~+~ \bar{\epsilon}^\ad \non \\
&& x'^{\a \ad} ~=~ x^{\a \ad} ~+~ a^{\a \ad}
~-~\frac{i}{2} \left( \epsilon^\a \thb^\ad ~+~
\bar{\epsilon}^\ad \th^\a \right)
\label{transf1}
\eea
is generated by two
complex charges $Q_\a$ ($\bar{Q}_\ad = Q_\a^\dag$) and the four--momentum
$P_{\a \ad}$ subjected to
\beq
\{ Q_\a , Q_\b \} ~=~ \{ \bar{Q}_\ad , \bar{Q}_\bd \} ~=~ 0  \quad , \quad
\{ Q_\a , \bar{Q}_\ad \} ~=~ P_{\a \ad}
\eeq

Representations of supersymmetry are given by superfields
$V(x^{\a \ad}, \th^\a, \thb^{\ad})$ whose components are obtained by expanding
$V$ in powers of the spinorial coordinates. The set of superfields is closed
under the standard product of functions. The product of two superfields
is (anti)commutative, $V \cdot W = (-1)^{deg(V) \cdot deg(W)} W \cdot V$,
and associative, $(K \cdot V) \cdot W = K \cdot (V \cdot W)$.

In order to define a non(anti)commutative superspace,
we consider the most
general structure of the algebra for a set of four bosonic real coordinates
and a complex two--component Weyl spinor with $(\th^\a)^\dag = \thb^\ad$
\bea
&&\left\{\th^{\a},\th^{\b}\right\} ~=~{\cal A}^{\a\b}(x,\th,
\bar{\th}) \qquad , \qquad
\left\{\bar{\th}^{\dot{\a}},\bar{\th}^{\dot{\b}}\right\} ~=~
\bar{\cal A}^{\dot{\a}\bd}(x,\th,\bar{\th})
\non\\
&&\left\{\th^{\a},\bar{\th}^{\dot{\a}}\right\}
~=~{\cal B}^{\a\dot{\a}}
(x,\th,\bar{\th})\non\\
&&\left[x^{\underline{a}},\th^{\b}\right]~=~i{\cal C}^{\underline{a}\b}
(x,\th,\bar{\th})\qquad , \qquad
\left[x^{\underline{a}},\bar{\th}^{\dot{\b}}\right]~=~
i\bar{\cal C}^{\underline{a} \dot{\b}}(x,\th,\bar{\th})\non \\
&&\left[x^{\underline{a}},x^{\underline{b}}\right]~=~
i{\cal D}^{\underline{a} \underline{b}}
(x,\th,\bar{\th})
\label{coord2}
\ena
Here, ${\cal A}, {\cal B}, {\cal C}, {\cal D}$ are local functions of the
superspace variables and we have defined $\bar{\cal A}^{\ad \bd}
\equiv ({\cal A}^{\a \b})^\dag$,
$\bar{\cal C}^{\underline{a} \bd} \equiv ({\cal C}^{\underline{a} \b})^\dag$.
From the conjugation rules for the coordinates it follows also
$\left({\cal B}^{\a\dot{\a}}\right)^{\dag}={\cal B}^{\a\dot{\a}}$ and
$\left({\cal D}^{\underline{a} \underline{b}}\right)^{\dag} =
{\cal D}^{\underline{a} \underline{b}}$.

To implement (\ref{coord2}) to be the algebra of the coordinates
of a non(anti)commutative $N=1$ superspace we require its covariance
under the group of space translations and supertranslations (\ref{transf1}).
The covariance under observer--Lorentz transformations is manifest in
(\ref{coord2}), while we do not require covariance under particle--Lorentz
transformations
which is in general broken in a noncommutative geometry (for a discussion
of the two different Lorentz transformations see \cite{lorentz}).
We restrict our analysis to the case of an
undeformed group where the parameters $a^{\a\ad}$, $\epsilon^\a$
and $\bar{\epsilon}^\ad$ in (\ref{transf1}) are kept
(anti)commuting \footnote{More general
constructions of non(anti)commutative geometries in grassmannian spaces
have been
considered, where also the algebra of the parameters is deformed
\cite{kosinski}.}.

Imposing covariance amounts to ask the functional dependence of the
${\cal A} , {\cal B}, {\cal C} , {\cal D}$ in (\ref{coord2}) to be the
same at {\em any} point of the supermanifold.
To work out explicitly the constraints which follow,
we perform a (super)translation (\ref{transf1}) on the coordinates
and compute the algebra of the new coordinates in terms of
the old ones.
We find that the functions appearing in (\ref{coord2}) are constrained by 
the following set of independent equations
\beq
{\cal A}^{\a \b} (x',\th',\thb' ) ~=~
{\cal A}^{\a \b} (x,\th,\thb ) \quad , \quad
{\cal B}^{\a \ad} (x',\th',\thb' ) ~=~
{\cal B}^{\a \ad} (x,\th,\thb )
\label{first}
\eeq
\beq
{\cal C}^{\a\dot{\a}\b}(x',\th',\thb') ~=~
{\cal C}^{\a\dot{\a}\b}(x,\th,\thb )~-~\frac{1}{2}
\e^{\a}{\cal B}^{\b \ad}(x,\th,\thb )~-~\frac{1}{2}
\bar{\e}^{\ad}{\cal A}^{\a\b}(x,\th,\thb )
\label{second}
\eeq
\bea
&& {\cal D}^{\a\dot{\a}\b\dot{\b}}(x',\th',\thb') ~=~
{\cal D}^{\a\dot{\a}\b\dot{\b}}(x,\th,\overline{\th})
\non\\
&&-~
\frac{i}{2}\left(\e^{\b}\bar{\cal C}^{\a\dot{\a}\dot{\b}}
(x,\th,\thb )
+\bar{\e}^{\dot{\b}}{\cal C}^{\a\dot{\a}\b}(x,\th,\thb)
-\e^{\a}\bar{\cal C}^{\b\dot{\b}\ad}(x,\th,\thb )
-\bar{\e}^{\dot{\a}}{\cal C}^{\b\dot{\b}\a}(x,\th,\thb)\right)
\non\\
&& -\frac{i}{4}\left(\e^{\a}\bar{\cal A}^{\dot{\a}\dot{\b}}
(x,\th,\thb )\e^{\b}~+~\e^{\a} {\cal B}^{\b\dot{\a}}
(x,\th,\thb )\bar{\e}^{\dot{\b}} \right.
\non\\
&& ~~~~~~~~~~~~~~~~~~~~~~~~~~~~~\left. +\bar{\e}^{\dot{\a}}
{\cal B}^{\a\dot{\b}}(x,\th,\thb )\e^{\b}+
\bar{\e}^{\dot{\a}}{\cal A}^{\a\b}(x,\th,\thb )
\bar{\e}^{\dot{\b}}\right)
\non\\
&&~~~~~~~~
\label{third}
\eea
together with their hermitian conjugates.

Looking for the most general local solution brings to the following algebra
for a non(anti)commutative
geometry in Minkowski superspace consistent with (super)translations
\bea
\left\{\th^{\a},\th^{\b}\right\} &&=~ A^{\a\b} \quad , \quad
\left\{ \thb^{\ad} , \thb^{\bd} \right\} ~=~ \bar{A}^{\ad \bd}
\quad , \quad
\left\{ \th^{\a},\thb^{\dot{\a}} \right\} ~=~ B^{\a\dot{\a}}
\non\\
\left[ x^{\a\dot{\a}},\th^{\b} \right] &&=~
i {\cal C}^{\a\dot{\a}\b} (\th, \thb)
\non\\
\left[x^{\a\dot{\a}},\thb^{\dot{\b}}\right] &&=~
i \bar{{\cal C}}^{\a\dot{\a}\dot{\b}}(\th, \thb)
\non\\
\left[x^{\a\dot{\a}},x^{\b\dot{\b}}\right]
&&=~ i {\cal D}^{\a\dot{\a}\b\dot{\b}}(\th,\thb)
\label{nonanti}
\eea
where
\bea
&& {\cal C}^{\a\dot{\a}\b}(\th,\thb ) ~=~
C^{\a\dot{\a}\b} ~-~\frac{1}{2}\th^{\a} B^{\b\dot{\a}}
~-~ \frac{1}{2}\thb^{\dot{\a}} A^{\a\b}\cr
&&{\cal D}^{\a\dot{\a}\b\dot{\b}}(\th,\thb )
=~ D^{\a\dot{\a}\b\dot{\b}}
~-~\frac{i}{2}\left(\th^{\b} \bar{C}^{\a \dot{\a}\dot{\b}}
~-~\thb^{\dot{\a}} C^{\b\dot{\b}\a}
~-~ \th^{\a} \bar{C}^{\b\dot{\b}\dot{\a}}
~+~ \thb^{\dot{\b}} C^{\a\dot{\a}\b}\right)
\non\\
&&-~\frac{i}{4}\left(\th^{\a} \bar{A}^{\dot{\a}\dot{\b}} \th^{\b}
~+~\th^{\a} B^{\b\dot{\a}} \thb^{\dot{\b}}
~+~ \thb^{\dot{\a}} B^{\a\dot{\b}}\th^{\b}
~+~ \thb^{\dot{\a}} A^{\a\b} \thb^{\dot{\b}} \right)
\label{choice}
\eea
and $A$, $B$, $C$ and $D$ are constant functions.

We notice that, while covariance under spacetime
translations necessarily requires the non(anti)commutation functions to be
independent of the $x$ coordinates, the covariance under supersymmetry is
less restrictive and allows for a particular dependence on the spinorial
coordinates.

On the algebra of smooth functions of the superspace variables we can formally
define a graded bracket which reproduces the fundamental
algebra (\ref{nonanti}) when applied to the coordinates.
In the case of bosonic Minkowski spacetime,
the noncommutative algebra (\ref{bosonic}) can be obtained by
interpreting the l.h.s. of this relation as the Poisson bracket of classical
commuting variables, where, for generic functions of spacetime, the Poisson
bracket is defined as $\{ f , g \}_{P} = i \Theta^{\m \n}\pa_\m f \pa_\n g$.
Generalizing to Minkowski superspace, the graded bracket
must be constructed as a bidifferential operator with respect to the
superspace variables. 
Using covariant derivatives $D_A \equiv (D_\a, \bar{D}_\ad,
\pa_{\a \ad})$, for generic functions $\Phi$ and $\Psi$ of the superspace
coordinates we define the bidifferential operator 
\beq
\{ \Phi , \Psi \}_P ~=~ \Phi \overleftarrow{D}_A \, P^{AB} \,
\overrightarrow{D}_B \Psi
\label{poisson1}
\eeq
where
\beq
P^{AB} ~\equiv~
\pmatrix{
P^{\a \b}  & P^{\a \bd} & P^{\a \underline{b}} \cr
P^{\ad \b} & P^{\ad \bd} & P^{\ad \underline{b}}  \cr
P^{\underline{a} \b} & P^{\underline{a} \bd} &
P^{\underline{a} \underline{b}} \cr }
~=~ \pmatrix{
-A^{\a \b}  & -B^{\a \bd} & iC^{\b \bd \a} \cr
-B^{\ad \b} & -\bar{A}^{\ad \bd} & i\bar{C}^{\b \bd \ad}  \cr
iC^{\a \ad \b} & i\bar{C}^{\a \ad \bd} & iD^{\a \ad \b \bd} \cr }
\label{supermatrix}
\eeq
is a constant graded symplectic supermatrix satisfying
$P^{BA} = (-1)^{(a+1)(b+1)} P^{AB}$, $a$ denoting the grading of $A$.
It is easy to verify that applying this
operator to the superspace coordinates we obtain (\ref{nonanti}).

Alternatively, one could express the graded brackets (\ref{poisson1}) 
in terms of torsion free, noncovariant spinorial derivatives
$(\pa_\a, \bar{\pa}_\ad, \pa_{\a \ad})$ so obtaining a matrix $P^{A B}$
explicitly dependent on $(\th, \thb)$.

\vskip 15pt
It is important to notice that the non(anti)commutative extension given in
(\ref{nonanti}) in general deforms the supersymmetry algebra. 
In the standard
case, defining $Q_A \equiv (Q_\a, \bar{Q}_\ad, -i\pa_{\a\ad})$, the
supersymmetry algebra can be written as
\bea
&& [Q_A,Q_B\} ~=~ i {T_{AB}}^C Q_C \quad , \quad
[D_A,D_B\} ~=~ {T_{AB}}^C D_C
\non\\
&& [Q_A,D_B\} ~=~ 0
\label{standard}
\eea
where ${T_{AB}}^C$ is the torsion of the flat superspace
(${T_{\a \bd}}^{\underline{c}} = {T_{\bd \a}}^{\underline{c}} = i \d^{~\g}_\a
\d_\bd^{~\dot{\g}}$ are the only nonzero components) and we have introduced
the notation $[F_A,G_B\} \equiv F_A G_B - (-1)^{ab} G_B F_A$.
Turning on non(anti)commutation in superspace leads instead to
\bea
&& [Q_A,Q_B\} ~=~ i {T_{AB}}^C Q_C ~+~ {R_{AB}}^{CD} Q_C Q_D
\non\\
&& [D_A,D_B\} ~=~ {T_{AB}}^C D_C ~+~ {R_{AB}}^{CD} D_C D_D
\non\\
&& [Q_A,D_B\} ~=~ {R_{AB}}^{CD} Q_C D_D
\label{modalg}
\eea
where ${T_{AB}}^C$ is still the torsion of the flat superspace, while
\beq
{R_{AB}}^{CD} ~=~ -\frac{1}{8} \, P^{MN} {T_{M[A}}^C {T_{B)N}}^D
\label{rtensor2}
\eeq
($[ab)$ means antisymmetrization when at least one of the indices
is a vector index, symmetrization otherwise)
is a curvature tensor whose presence is a direct consequence
of the non(anti)commutation of the grassmannian coordinates.
Its nonvanishing components are
\bea
&& {R_{\a \b}}^{\underline{c} \underline{d}} ~=~ \frac{1}{8}
P^{\dot{\gamma} \dot{\delta}} \d_{(\a}^{~\g} \d_{\b)}^{~\d}
\quad , \quad
{R_{\ad \bd}}^{\underline{c} \underline{d}} ~=~ \frac{1}{8}
P^{\gamma \delta} \d_{(\ad}^{~\dot{\g}} \d_{\bd)}^{~\dot{\d}}
\non\\
&& {R_{\a \bd}}^{\underline{c} \underline{d}} ~=~
{R_{\bd \a}}^{\underline{c} \underline{d}} ~=~ \frac{1}{8}
\left( P^{\gamma \dot{\d}} \d_\a^{~\d} \d_\bd^{~\dot{\g}}
~+~ P^{\d \dot{\g}} \d_\a^{~\g} \d_\bd^{~\dot{\d}} \right)
\label{rtensor}
\eea

Since the terms proportional to the curvature in the algebra (\ref{modalg})
are quadratic in the supersymmetry charges and covariant derivatives,
we can define new graded brackets
\bea
&& [ Q_A , Q_B \}_q ~\equiv~ Q_A Q_B ~-~ (-1)^{ab} [ {\d_B}^C {\d_A}^D ~+~
(-1)^{ab} {R_{AB}}^{CD} ] Q_C Q_D
\eea
and analogous ones for $[ D_A , D_B \}_q$ and $[ Q_A , D_B \}_q$, 
which satisfy the standard algebra (\ref{standard}).
The new brackets can be interpreted as a quantum deformation
associated to a $q$--parameter which in this case is a rank--four tensor
\beq
{q_{AB}}^{CD} ~\equiv~ {\d_B}^C {\d_A}^D ~+~ (-1)^{ab} {R_{AB}}^{CD}
\eeq
It would be interesting to investigate this issue further.

\sect{Discussing associativity}

Given the bidifferential operator (\ref{poisson1}) associated to 
the noncommutative supergeometry defined in (\ref{nonanti}) it is easy to prove
the following identities
\bea
&& \{ \Phi, \Psi \}_P ~=~ (-1)^{1+ deg(\Phi) \cdot deg(\Psi)} \, \{ \Psi ,
\Phi \}_P
\non \\
&& \{ c \Phi ,\Psi \}_P ~=~ c \, \{ \Phi, \Psi \}_P  \quad , \quad
\{ \Phi ,c \Psi \}_P ~=~ (-1)^{deg(c) \cdot deg(\Phi)} \,c \, \{ \Phi,
\Psi \}_P
\non\\
&& \{ \Phi + \Psi , \Omega \}_P ~=~ \{ \Phi , \Omega \}_P ~+~ \{\Psi,
\Omega\}_P
\eea
The operator $\{ \, , \}_P$ will then be promoted to a 
graded Poisson structure
on superspace if and only if the Jacobi identities hold
\bea
&& \{ \Phi, \{ \Psi , \Omega \}_P \}_P +
(-1)^{deg(\Phi)\cdot [deg(\Psi) + deg(\Omega)]}
\{ \Psi, \{ \Phi, \Omega \}_P \}_P
\non\\
&& ~~~~~~~~~~~~~~~~~~~~~~+ (-1)^{deg(\Omega)\cdot [deg(\Phi) + deg(\Psi)]}
\{ \Omega, \{ \Phi, \Psi \}_P \}_P
~=~ 0
\label{jacobi}
\eea
for any triplet of functions of the superspace variables.
This property is not in general satisfied
as a consequence of possible lack of
associativity in the fundamental algebra (\ref{nonanti}). 
Indeed, imposing (\ref{jacobi}) yields the nontrivial
conditions
\begin{eqnarray}
P^{AR} P^{BS}T_{SR}^{\;\;\;\;C}(-1)^{c+b(c+a+r)} &+&
P^{BR}P^{CS}T_{SR}^{\;\;\;\;A}(-1)^{a+c(a+b+r)} \nonumber \\
&+& \, P^{CR}P^{AS}T_{SR}^{\;\;\;\;B}(-1)^{b+a(b+c+r)} = 0
\label{Jac_cov}
\end{eqnarray}
\beq
(-1)^{bm} P^{AM} P^{BN} {R_{MN}}^{CD} ~=~ 0
\label{Jac_cov2}
\eeq
where the torsion ${T_{AB}}^C$ and the curvature ${R_{AB}}^{CD}$ have been
introduced in (\ref{standard}) and (\ref{modalg}).

If $P^{AB}$ is invertible ($P_{AB}P^{BC}=\delta_A^C$), 
the equation (\ref{Jac_cov}) is equivalent to
the vanishing of the contorsion tensor $H_{ABC}$ defined by
\begin{equation}
H_{ABC} = T_{AB}^{\;\;\;\;D}P_{DC}(-1)^{ac} +
T_{CA}^{\;\;\;\;D}P_{DB}(-1)^{cb} + T_{BC}^{\;\;\;\;D}P_{DA}(-1)^{ba}\,.
\end{equation}
Its only nonvanishing components are 
\bea
&& H_{\a \dot{\a} \b}=-i\left[ P_{\a \dot{\a}~ \b} + P_{\b \dot{\a}~ \a}
\right]
\non\\
&&H_{\a \dot{\a} \dot{\b}}=-i \left[ P_{\a\dot{\a}~\dot{\b}} +
P_{\a\dot{\b}~\dot{\a}} \right] 
\non\\
&&H_{\a \dot{\a} \underline{b}}=iP_{\a \dot{\a} ~\underline{b}}
\eea
We notice that its bosonic components $H_{\underline{a} \underline{b} 
\underline{c}}$ vanish due to the $x$--independence of the noncommutation
functions in (\ref{nonanti}). The nonvanishing of $H$ comes entirely from
the $\theta$--dependence of the functions in (\ref{choice}).

In a string theory context, knowing the geometric objects of the 
(super)space like the curvature $R$ and the field strength $H$ would
allow to identify  the class of supergravity backgrounds in which
noncommutative geometries might naturally emerge.\\
In \cite{Cornalba:2001sm} the deformation of D-brane world-volumes
in the presence of NS-NS curved backgrounds was investigated in a 
nonsupersymmetric context.
It was shown that, if both the brane and the background are curved, i.e.
$H \equiv d(B +F) \neq 0$, then
the deformation of the world-volume is a Kontsevich deformation
which defines a nonassociative, noncommutative product.
Noncommutativity is governed by the usual NS-NS $B$-field,
whereas nonassociativity arises from the NS-NS field strength
$H$.
This suggests that our supergeometries might naturally appear, if an
analysis similar to \cite{Cornalba:2001sm} were to be performed in a
manifestly supersymmetric formalism
(i.e. working with a Green-Schwarz \cite{GrSchw} or
Berkovits \cite{Berkovits:2000fe} string) for backgrounds with 
nonvanishing super p--form field strengths.
First steps in this direction were taken in \cite{chuzamora}, where open
Green-Schwarz superstrings ending
on a D-brane in the presence of a constant NS-NS $B$ field in flat
spacetime were considered in a manifestly supersymmetric approach.
In this simple case it was found that the anticommutation relations
for the fermionic
variables of superspace remain unmodified. It would be interesting to
extend these calculations to more general backgrounds to see whether
also a deformation of superspace can arise. Since the
noncommutative geometries we have constructed are characterized by a 3--form
field strength with the only nonvanishing components being fermionic, we
expect to find connections even with string backgrounds with 
$B_{\underline{a} \underline{b}}=0$.\\

We now search for the most general solutions of the conditions 
(\ref{Jac_cov}, \ref{Jac_cov2}).
Writing them in terms of the $P^{AB}$ components (\ref{supermatrix}) we obtain
\bea
&& B^{\a\dot{\b}} A^{\b\g} ~+~ A^{\b\a}B^{\g\dot{\b}} ~=~ 0
\non\\
&& B^{\a\dot{\b}}B^{\b\dot{\g}} ~+~ A^{\b\a} \bar{A}^{\dot{\b}\dot{\g}}
~=~ 0
\non\\
&& \left( \bar{C}^{\b\dot{\b}\dot{\a}} A^{\a\g}
~+~ C^{\b\dot{\b}\a} B^{\g\dot{\a}} ~-~
\bar{C}^{\a\dot{\a}\dot{\b}} A^{\b\g}
~-~ C^{\a\dot{\a}\b} B^{\g\dot{\b}} \right) ~=~ 0
\non\\
&& {\cal I}{\rm m}
\left( \bar{C}^{\a\dot{\a}\dot{\b}} C^{\g\dot{\g}\b} ~+~
\bar{C}^{\g\dot{\g}\dot{\a}} C^{\b\dot{\b}\a}
~+~ \bar{C}^{\b\dot{\b}\dot{\g}} C^{\a\dot{\a}\g}\right) ~=~ 0\,.
\label{jacobi4}
\eea
The first two conditions necessarily imply the vanishing
of the constants $A$ and $B$.
Inserting this result in the third constraint we immediately realize
that it is automatically satisfied and the only nontrivial condition 
which survives is the last one. 
This equation has nontrivial solutions. For example, the matrix
\beq
C^{\a \ad \b} ~=~
\pmatrix{
\psi^\b & \psi^\b \cr
\psi^\b & \psi^\b \cr }
\eeq
for any spinor $\psi^\b$, is a solution.
It would correspond to assume the
same commutations rules among any bosonic coordinate and the spinorial
variables.

We conclude that the most general {\em associative} and non(anti)commuting
algebra in Minkowski superspace has the form
\bea
\left\{\th^{\a},\th^{\b}\right\} &&=~
\left\{ \thb^{\ad} , \thb^{\bd} \right\}
~=~ \left\{ \th^{\a},\thb^{\dot{\b}} \right\} ~=~ 0
\non\\
\left[ x^{\a\dot{\a}},\th^{\b} \right] &&=~ i C^{\a\dot{\a}\b}
\non\\
\left[x^{\a\dot{\a}},\thb^{\dot{\b}}\right] &&=~ i
\bar{C}^{\a\dot{\a}\dot{\b}} \label{assnonanti} \\
\left[x^{\a\dot{\a}},x^{\b\dot{\b}}\right] &&=~
i D^{\a\dot{\a}\b\dot{\b}}
~+~\frac{1}{2} \left( \bar{C}^{\b\dot{\b}\dot{\a}}
\th^{\a} ~-~ \bar{C}^{\a\dot{\a}\dot{\b}} \th^\b
~+~ C^{\b\dot{\b}\a} \thb^{\dot{\a}}
~-~ C^{\a\dot{\a}\b} \thb^\bd \right)\,, \nonumber
\eea
where $C$ is subject to the last constraint in (\ref{jacobi4}).
Setting $C^{\a \ad \b}=0$ we recover the usual noncommuting
superspace considered so far in literature \cite{ferrara,chuzamora,all}.

Under conditions (\ref{jacobi4}) the graded brackets 
(\ref{poisson1}) satisfy the Jacobi identities (\ref{jacobi}), as can be easily
proved by expanding the functions in power series.
In this case we have a well--defined super Poisson structure on superspace.

We notice that a non(anti)commutative but associative geometry always
mantains the standard algebra (\ref{standard}) for the covariant derivatives. 
In fact,
in this case, from (\ref{rtensor}) it follows ${R_{AB}}^{CD} =0$.

\sect{Towards the construction of a noncommutative product}

In this section we describe the first few steps towards the construction
of a star product defined on the class of general superfields.
By definition, this product must be associative, i.e. it has 
to satisfy the Jacobi identities (\ref{jacobi}) when the fundamental
algebra is associative. 

In the 
nonsupersymmetric case the lack of associativity of the fundamental algebra
is signaled by the presence of a nonvanishing 3--form $H$.
A product has been constructed \cite{kontsevich} so that the terms
violating the Jacobi identities are proportional to $H$. 
The product is then automatically associative when 
the fundamental algebra is.

In the present case we have shown that the lack of associativity 
in superspace is related to a nonvanishing super 3--form.  
This suggests the possibility to construct 
a super star product by suitably generalizing the Kontsevich 
construction \cite{kontsevich} to superspace.  

We begin by considering the Moyal--deformed product defined
in the usual way
\begin{equation}
\Phi \ast \Psi \equiv \Phi \exp(\hbar\overleftarrow{D}_A P^{AB}
                       \overrightarrow{D}_B) \Psi, \label{deformprod}
\end{equation}
where $\Phi$ and $\Psi$ are arbitrary superfields, and
$\hbar$ denotes a deformation parameter. In general, due to
the lack of (anti)commutativity among covariant derivatives (see eq.
(\ref{modalg})), it is easy to prove that the $\ast$--product is
not associative even when the Poisson brackets are.
However, inspired by the Kontsevich procedure \cite{kontsevich}, 
we perturbatively define a modified product $\star$ with the 
property to
be associative up to second order in $\hbar$ when the Jacobi identities
are satisfied. Precisely, we find an explicit form for the product
by imposing the Jacobi identities (\ref{jacobi})
to be violated at this order only by terms proportional to $H$. 
To this end we define
\begin{eqnarray}
\Phi \star \Psi &\equiv & \Phi\Psi ~+~ \hbar \Phi\overleftarrow{D}_A P^{AB}
\overrightarrow{D}_B \Psi ~+~ \frac{\hbar^2}{2}\Phi(\overleftarrow{D}_A P^{AB}
\overrightarrow{D}_B)(\overleftarrow{D}_C P^{CD}\overrightarrow{D}_D) \Psi
\nonumber \\
&& -\frac{\hbar^2}{3}\left(\overrightarrow{D}_A \Phi \, {\cal M}^{ABC}
\, \overrightarrow{D}_B\overrightarrow{D}_C \Psi - (-1)^c
\overrightarrow{D}_C\overrightarrow{D}_A \Phi \, {\cal M}^{ABC} \,
\overrightarrow{D}_B \Psi\right) \nonumber \\
&& + {\cal O}(\hbar^3)\,, \label{konts}
\end{eqnarray}
where
\begin{eqnarray}
{\cal M}^{ABC} &=& P^{AD}{T_{DE}}^C P^{EB} (-1)^{ce} +
            \frac12 P^{BD}{T_{DE}}^A P^{EC} (-1)^{ae + a + b + ab + bc}
            \nonumber \\
        & & + \frac12 P^{CD}{T_{DE}}^B P^{EA} (-1)^{be + a + c + ac + ab}.
\end{eqnarray}
Since it is straightforward to show that
\begin{eqnarray}
\lefteqn{(\Phi \star \Psi)\star \Omega ~-~ \Phi \star (\Psi \star \Omega)
~=~}
\nonumber \\
& & -\frac 23 \hbar^2(-1)^{(c+b)(e+1)+eg+cf}\overrightarrow{D}_A \Phi \,
    P^{AE}P^{BF}P^{CG} \, H_{GFE} \, \overrightarrow{D}_C \Omega \,
    \overrightarrow{D}_B \Psi \\
& & ~+~ {\cal O}(\hbar^3). \nonumber
\end{eqnarray}
up to second order in $\hbar$ the product is associative if and only if $H=0$,
i.e. the fundamental algebra is associative. 
We note that at this order only the contorsion enters the breaking
of associativity, being the curvature tensor $R$ of order $\hbar$.

It would be interesting to pursue the construction of the star product to
all orders in $\hbar$. We believe that to this respect there are no objections 
of principle in generalizing to superspace 
the Kontsevich procedure to all orders. 

We now discuss the closure of the class of chiral superfields under the
deformed products we have introduced. For a generic choice
of the supermatrix $P^{AB}$ the star product of two chiral superfields
(satisfying $\bar{D}_{\dot{\a}}\Phi = 0$)
is {\em not} a chiral superfield, both for associative and nonassociative
products.
However, in the particular case where the only nonvanishing components of
the symplectic supermatrix $P^{AB}$ are $P^{\a\dot{\b}}$ and
$P^{\underline{a}\underline{b}}$, chiral superfields
are closed both under
the deformed product defined in (\ref{deformprod}) and under the Kontsevich
star product (\ref{konts}) (for the latter up to terms of order
${\cal O}(\hbar^3)$).
Clearly for $P^{\a\dot{\b}} \neq 0$ the above star products
are no more associative.
For chiral superfields however, they become commutative\footnote{Generalized
star products that are commutative but nonassociative have been considered
in a different context in \cite{dastrivedi}.}.
This commutativity implies that there is no ambiguity in putting
the parenthesis e.~g.~in the cubic interaction term of a deformed
Wess--Zumino model, with action
\begin{equation}
S = \int d^4x d^2\theta d^2\bar{\theta}\,\Phi\star\bar{\Phi}
+ \int d^4x \left[\int d^2\theta \left(\frac m2 \Phi\star\Phi + \frac g3
\Phi\star\Phi\star\Phi\right) + {\mbox{c.~c.}} \right].
\label{WZ}
\end{equation}
Notice that in this case the $\star$--product in the kinetic action cannot
be simply substituted with the standard product as it happens in superspace
geometries where grassmannian coordinates anticommute \cite{ferrara, all}.

\sect{Non(anti)commutative $N=2$ Euclidean superspace}

The main difference in the description of euclidean superspace
with respect to Minkowski relies on the reality conditions
satisfied by the spinorial variables.
As it is well known \cite{PVP}, in euclidean signature
a reality condition on spinors is applicable only in the presence of
extended supersymmetry. We concentrate on
the simplest case, the $N=2$ euclidean superspace even if our analysis
can be easily extended to more general cases. In a chiral description
the two--component Weyl spinors satisfy a symplectic Majorana condition
\beq
(\th^\a_i )^{\ast} ~=~ \th_\a^i ~\equiv~ C^{ij} \, \th^\b_j \, C_{\b \a}
\quad , \quad
(\thb^{\ad, i} )^{\ast} ~=~ \thb_{\ad, i} ~\equiv~
\thb^{\bd, j} \, C_{\bd \ad} \, C_{ji}
\eeq
where $C^{12} = -C_{12} = i$.
This implies that the most general non(anti)commutative
algebra can be written as an obvious generalization of (\ref{coord2})
with the functions on the rhs now being in suitable representations
of the R--symmetry group. 
When imposing covariance under (super)translations we obtain
that the most general non(anti)commutative geometry in euclidean
superspace is
\bea
\left\{\th^{\a}_i,\th^{\b}_j\right\} &&=~ {A_1}^{\a\b,}_{~~~ij} \quad , \quad
\left\{ \thb^{\ad,i} , \thb^{\bd,j} \right\} ~=~ A_2^{\ad \bd,ij}
\quad , \quad
\left\{ \th^{\a}_i,\thb^{\dot{\a},j} \right\} ~=~ B^{\a\dot{\a},~~j}_{~~~~i}
\non\\
\left[ x^{\a\dot{\a}},\th^{\b}_i \right] &&=~
i{{\cal C}_1}^{\underline{a}\b,}_{~~~ i} (\th, \thb)
\non\\
\left[x^{\a\dot{\a}},\thb^{\dot{\b},i}\right] &&=~
i {\cal C}_2^{\a\dot{\a}\dot{\b},i}(\th, \thb)
\non\\
\left[x^{\a\dot{\a}},x^{\b\dot{\b}}\right] &&=~
i {\cal D}^{\a\dot{\a}\b\dot{\b}}(\th,\thb)
\label{nonanti2}
\eea
where 
\bea
&& {{\cal C}_1}^{\a\dot{\a}\b,}_{~~~~~i} (\th, \thb) ~\equiv~
{C_1}^{\a\dot{\a}\b,}_{~~~~~i}
~+~\frac{i}{2} \th^{\a}_j B^{\b \dot{\a},~~j}_{~~~~i}
~+~ \frac{i}{2} \thb^{\dot{\a}, j} {A_1}^{\a\b,}_{~~~ji}
\non\\
&& {\cal C}_2^{\a\dot{\a}\dot{\b},i}(\th, \thb) ~\equiv~
C_2^{\a\dot{\a}\dot{\b},i}
~+~\frac{i}{2} \th^{\a}_j A_2^{\ad \bd,ji}
~+~\frac{i}{2} \thb^{\ad, j} B^{\a\dot{\b},~~i}_{~~~~j}
\non\\
&& {\cal D}^{\a\dot{\a}\b\dot{\b}}(\th,\thb) ~\equiv~
D^{\a\dot{\a}\b\dot{\b}}
\non\\
&&~~~~~~~
+~\frac{1}{2} \left( \th^{\a}_i C_2^{\b\dot{\b}\dot{\a},i}
~-~ \th^\b_i C_2^{\a\dot{\a}\dot{\b},i}
~+~ \thb^{\dot{\a}, i} {C_1}^{\b\bd\a,}_{~~~~~ i}
~-~ \thb^{\bd,i}  {C_1}^{\a\ad\b,}_{~~~~~ i} \right)
\non\\
&&~~~~~~~
+~\frac{i}{4}\left( \th^{\a}_i \, A_2^{\dot{\a}\dot{\b},ij} \, \th^{\b}_j
~+~ \th^{\a}_i \, B^{\b \dot{\a},~~i}_{~~~~j} \, \thb^{\dot{\b},j}
~+~\thb^{\dot{\a},i} \, B^{\a \dot{\b},~~j}_{~~~~i} \, \th^{\b}_j
~+~\thb^{\dot{\a},i} \, {A_1}^{\a\b,}_{~~~ij} \, \thb^{\dot{\b},j} \right)
\non\\
&&~~~~~~~~~
\label{cal2}
\eea
with $A_1$, $A_2$, $B$, $C_1$, $C_2$ and $D$ constant.

Following the same steps as in the Minkowski case, we can look
for the most general associative algebra. The results we obtain for
{\em associative} non(anti)commuting geometries in euclidean superspace are
\bea
\left\{\th^{\a}_i ,\th^{\b}_j \right\} &&=~ {A_1}^{\a\b,}_{~~~ij} \quad ,\quad
\left\{ \thb^{\ad,i} , \thb^{\bd,j} \right\} ~=~ 0 \quad , \quad
\left\{ \th^{\a}_i,\thb^{\dot{\b},j} \right\} ~=~ 0
\non\\
\left[ x^{\a\dot{\a}},\th^{\b}_i \right] &&=~ i{C_1}^{\a\dot{\a}\b,}_{~~~~~i}
~-~ \frac12 \thb^{\ad,j} {A_1}^{\a \b,}_{~~~ji}
\non\\
\left[x^{\a\dot{\a}},\thb^{\dot{\b},i}\right] &&=~ 0
\\
\left[x^{\a\dot{\a}},x^{\b\dot{\b}}\right] &&=~
iD^{\a\dot{\a}\b\dot{\b}}
~+~ \frac{i}{2} \left(\thb^{\dot{\a},i}  {C_1}^{\b\dot{\b}\a,}_{~~~~~i}
~-~ \thb^{\bd,i} {C_1}^{\a\dot{\a}\b,}_{~~~~~i}  \right) ~-~
\frac14 \thb^{\ad,i} \, {A_1}^{\a \b,}_{~~~ij} \, \thb^{\bd,j}
\non
\label{assnonantie1}
\eea
or
\bea
\left\{\th^{\a}_i,\th^{\b}_j\right\} &&=~ 0 \quad , \quad
\left\{ \thb^{\ad,i} , \thb^{\bd,j} \right\}
~=~ A_2^{\ad \bd,ij} \quad ,\quad
\left\{ \th^{\a}_i,\thb^{\dot{\b},j} \right\} ~=~ 0
\non\\
\left[ x^{\a\dot{\a}},\th^{\b}_i \right] &&=~ 0
\non\\
\left[x^{\a\dot{\a}},\thb^{\dot{\b},i}\right] &&=~ iC_2^{\a\dot{\a}\dot{\b},i}
~-~ \frac12 \th^\a_j A_2^{\ad \bd , ji}
\\
\left[x^{\a\dot{\a}},x^{\b\dot{\b}}\right] &&=~
iD^{\a\dot{\a}\b\dot{\b}}
~+~\frac{i}{2} \left( \th^{\a}_i C_2^{\b\dot{\b}\dot{\a},i}
~-~ \th^\b_i C_2^{\a\dot{\a}\dot{\b},i}  \right)
~-~ \frac14 \th^\a_i \, A_2^{\ad \bd , ij} \, \th^\b_j
\non
\label{assnonantie2}
\eea
We notice that in this case associativity imposes less restrictive constraints
because of the absence of conjugation relations between $A_1$ and $A_2$. 
As a consequence, nontrivial anticommutation relations among $\theta$'s 
(or $\bar{\theta}$'s) are allowed.
Moreover the R--symmetry group of the $N=2$ euclidean superalgebra is
broken only by the constant terms $C_1$ and $C_2$. Setting these terms equal
to zero leads to nontrivial (anti)commutation relations preserving 
$R$--symmetry.

Again, explicit expressions for the corresponding graded brackets
can be obtained as an obvious generalization of (\ref{poisson1}--
\ref{supermatrix}). In this case
they define a super Poisson structure on the euclidean superspace.
A simple example of a super Poisson structure is
\beq
\{ \Phi , \Psi \}_P ~=~ - ~\Phi \overleftarrow{D}_{\a}^i \,
{A_1}^{\a \b,}_{~~~ij} \, \overrightarrow{D}_{\b}^j \Psi
\label{example2}
\eeq
We notice that this extension is allowed {\em only} in euclidean superspace,
where it is consistent with the reality conditions on the spinorial
variables.

\section{Final remarks}

In this paper we have studied the most general non(anti)commutative
geometry in $N=1$ four dimensional Minkowski superspace that is compatible
with classical (super)translations.
We have shown that nonanticommutation relations
among spinorial variables are allowed if the commutation relations of
bosonic coordinates with the spinorial ones and bosonic coordinates among
themselves acquire a particular dependence on the $\th$--variables.
In a geometric framework we can interpret the supermatrix $P^{AB}$
defining the non(anti)commutative algebra (see eq. (\ref{supermatrix})) as a
nontrivial metric in superspace. The geometry is in general nonassociative
and deforms the algebra of the superspace derivatives
through a curvature term (which, however, does not affect the algebra
of the coordinates). This
deformation can be interpreted as a quantum deformation associated to
a four--rank tensor $q$. The geometric properties of the $q$--deformed
superspace haven't been considered in this paper but certainly deserve a
deeper investigation.

We have further showed that imposing associativity, i.e. the validity
of the Jacobi identities,
implies additional conditions on the (anti)com\-mutat\-ors
of superspace coordinates which nevertheless allow for nontrivial
deformations involving also fermionic variables. In particular, the spinorial
variables are required to anticommute, but they can have nonzero commutation
relations with the bosonic coordinates.
As a consequence, in the associative case
the algebra of the (super)derivatives is not $q$--deformed.
Inspired by the Kontsevich procedure, we obtained the first three terms 
in a series expansion in the deformation parameter $\hbar$
for a possible noncommutative product that is associative up to second order, 
if the Jacobi identities are satisfied.
For the general case, the deviation from associativity
has been shown to be proportional to a super
three--form field strength which, in a complete covariant formalism, has
the interpretation of the contorsion of the non(anti)commutative superspace.
For the bosonic case, it was shown 
in \cite{Cornalba:2001sm} that the deformation of D--brane world-volumes in 
curved backgrounds,
with $H = d(B+F) \neq 0$, is described by a nonassociative 
Kontsevich star--product. It would be interesting to see if an analogous 
procedure 
performed in a manifestly supersymmetric formalism (Green--Schwarz or 
Berkovits) leads to superspace deformations involving fermionic 
coordinates.

We have extended our analysis to the case of $N=2$ euclidean
superspace. Due to the different hermiticity conditions on the spinorial
variables, the non(anti)commutat\-ive euclidean superspace manifests quite
different features from Minkowski. In fact, in this case nonzero
anticommutation relations among spinorial variables are allowed by
covariance and associativity. This implies that, even in the case of
non(anti)com\-mutat\-ive but associative geometry, the supersymmetry algebra
is $q$--deformed. The results obtained for the $N=2$ case can be easily
extended to generic $N >1$. It is however important to stress that in general
the R--symmetry group is broken by the noncommutativity among fermionic
and bosonic coordinates.

An interesting continuation of our work would be to
generalize the results of \cite{chuzamora}, in order to see
whether deformations of fermionic variables can arise for open
superstrings ending on a D--brane in the presence of a more general super 
field strength.
It would be also interesting to study field theories defined on a
non(anti)commutative superspace, such as the extended Wess--Zumino model
proposed in (\ref{WZ}). As already noticed, when the spinorial variables
satisfy a nontrivial algebra, also the kinetic action contains interaction
terms.

\end{document}